\documentclass[a4paper,fleqn]{cas-dc}
\usepackage[numbers]{natbib}

\usepackage{cancel}

\usepackage{amsmath}
\usepackage{amssymb}
\usepackage{subfigure} 
\usepackage{algorithm}
\usepackage{algorithmic}
\usepackage{multirow}
\usepackage{float}

\usepackage{hyperref}
\usepackage{fontawesome5}

\usepackage[justification=centering]{caption}
\usepackage{caption}

\def\tsc#1{\csdef{#1}{\textsc{\lowercase{#1}}\xspace}}
\tsc{WGM}
\tsc{QE}
\tsc{EP}
\tsc{PMS}
\tsc{BEC}
\tsc{DE}

\begin{document}
\let\WriteBookmarks\relax
\def\floatpagepagefraction{1}
\def\textpagefraction{.001}
\shorttitle{Data-driven multi-moment fluid modeling of Landau damping}
\shortauthors{Wenjie~Cheng et~al.}

\title [mode = title]{Data-driven, multi-moment fluid modeling of Landau damping}                      

\author[1,2]{Wenjie~Cheng} 
\author[1,2]{Haiyang~Fu{\footnotesize{{\href{mailto:haiyang_fu@fudan.edu.cn}{\faIcon{envelope}}}}}}
[style=chinese,orcid=0000-0003-1816-1024]
\cormark[1]

\author[3,4]{Liang~Wang}

\author[4,3]{Chuanfei~Dong{\footnotesize{{\href{mailto:dcfy@princeton.edu}{\faIcon{envelope}}}}}}
[style=chinese,orcid=0000-0002-8990-094X]
\cormark[1]

\author[1,2]{Yaqiu~Jin} 
\author[1,2]{Mingle~Jiang} 
\author[1,2]{Jiayu~Ma}
\author[1,2]{Yilan~Qin} 
\author[1,2]{Kexin~Liu}

\address[1]{School of Information Science and Engineering, Fudan University, Shanghai, 200433, China}
\address[2]{Key Laboratory for Information Science of Electromagnetic Waves (MoE), Fudan University, Shanghai, 200433, China}
\address[3]{Department of Astrophysical Sciences, Princeton University, Princeton, New Jersey, 08544, USA}
\address[4]{Princeton Plasma Physics Laboratory, Princeton University, Princeton, New Jersey, 08540, USA}

\cortext[cor1]{Corresponding authors}

\begin{abstract}
Deriving governing equations of complex physical systems based on first principles can be quite challenging when there are certain unknown terms and hidden physical mechanisms in the systems. In this work, we apply a deep learning architecture to learn fluid partial differential equations (PDEs) of a plasma system based on the data acquired from a fully kinetic model. The learned multi-moment fluid PDEs are demonstrated to incorporate kinetic effects such as Landau damping. Based on the learned fluid closure, the data-driven, multi-moment fluid modeling can well reproduce all the physical quantities derived from the fully kinetic model. The calculated damping rate of Landau damping is consistent with both the fully kinetic simulation and the linear theory. The data-driven fluid modeling of PDEs for complex physical systems may be applied to improve fluid closure and reduce the computational cost of multi-scale modeling of global systems. 
\end{abstract}


\maketitle

\section{\label{Introduction}Introduction}

A plasma system can be described by either microscopic kinetic theory that involves the dynamics of individual particles or macroscopic fluid equations. The kinetic theory can describe the distribution of particles in phase space. Although the kinetic model is accurate, the computational cost is high for large-scale problems. Most fluid models assume that the velocity distribution of particles obeys the Maxwellian distribution, which has a low computational cost with macroscopic averaged quantities such as density, velocity, and pressure from kinetic models. However, kinetic effects are not negligible in certain plasma problems such as ion-temperature-gradient instability \cite{Hammett1990}, collisionless magnetic reconnection \cite{Jonathan2020} and ionospheric modification experiment \cite{Fu2018,Fu2020}. Therefore, it is desired that macroscopic fluid models can integrate kinetic physics to resemble the kinetic model but with a cheap computational cost especially for global systems \cite{Wang2015,Dong2016,Wang2018,Dong2019,Jarmak2020,Wang2020,Rulke2021}. 

With progress in data-driven modeling of complex dynamical systems during the past decade, it is possible to extract physical laws and partial differential equations (PDEs) from real data. \emph{Schmidt et al.} \cite{schmidt2009distilling} proposed distilling natural laws from data using evolutionary symbolic regression to discover analytic relations automatically from motion tracking data without any physical priors. Symbolic regression based on neural networks has also been investigated for multiple physical data sets \cite{2020AI1,2020AI2,2020Integration}. 
Then, \emph{Brunton et al.} \cite{brunton2016discovering} and \emph{Rudy et al.} \cite{rudy2017data} proposed a sparse regression method PDE-FIND to discover the governing physical PDEs via Pareto analysis from spatiotemporal data. Recently, \emph{Raissi et al.} \cite{raissi2017physics,raissi2018hidden} proposed a physics-informed neural network (PINN) that integrates the physical model to solve and learn Navier Stokes fluid equations, which are applied to retrieve velocity and pressure field from density data \cite{raissi2020hfm}. \emph{Long et al.} \cite{long2017pde,long2019pde} proposed the PDE-NET method, which approximates the spatial differentiation through the convolutional neural network, and can realize long-time prediction even in noisy data. For electromagnetic problems, \emph{Xiong et al.} \cite{Xiong2019} and \emph{Fu et al.} \cite{Cheng2020} combined the sparse regression with the PDE-FIND to build a time network for learning Maxwell's equations and electromagnetic wave equations, respectively.

It has been pursued that accurate fluid models with kinetic physics can be obtained. The earliest work could be traced back to Grad's work \cite{Grad1949}. Based on Grad's work, \emph{Henning et al.} \cite{Henning2005} derived the well-known 13-moment equations by truncating the third-order distribution function in Hermite polynomials. Recently, machine learning has been adopted to learn multi-moment fluid models from kinetic data. \emph{Han et al.} \cite{Han2019} learned fluid closure for the hydrodynamic equations to approximate the kinetic model with high accuracy in the absence of scale separation. \emph{Zhang et al.} \cite{Zhang2020} adopts the PDE-FIND algorithm to learn the governing equations of fluid dynamics from simulation data of the Direct Simulation Monte Carlo (DSMC) method. Distinct from learning explicit PDEs as mentioned above, \emph{Ma et al.} \cite{Ma2020} applied machine learning to learn the surrogate models for the well-known Hammett-Perkins(H-P) closure \cite{Hammett1990} in plasma physics. In addition, \emph{Wang et al.} \cite{Wang2020Deep} also accurately reproduces the Landau damping rate and is more accurate than H-P closure, and \emph{Maulik et al.} \cite{Maulik2020Neural} reproduced the crucial physics inherent in known magnetized plasma closures. Thus, machine learning has shown promising potential to learn physical PDEs and accurate models from data.

Although the Hammett-Perkins (H-P) closure is simple in Fourier space, it is numerically challenging to implement in fluid simulation in configuration space. The closure by neural networks learned in the Ref. \cite{Ma2020} \cite{Wang2020Deep} \cite{Maulik2020Neural} are the black box that can not be physically interpretable. The limitation in Ref. \cite{Ma2020} is that only closure has been trained, and the training data comes from the well-known Hammett-Perkins (H-P) closure itself.

The purpose of our work is motivated by discovering the whole multi-moment fluid model PDEs directly from kinetic data based on machine learning. The closure form can not only be physically interpretable to capture kinetic effects but also is expected to be easily applied in configuration space for large-scale fluid simulations. Therefore, we attempt to construct a data-driven neural network architecture called \emph{mPDE-Net (Multiple Partial Differential Equations - Network)} to learn multi-moment fluid PDEs with closure from kinetic data. The \emph{mPDE-Net} method is extended to learn multiple-moment fluid model based on the work of \emph{Xiong et al.} \cite{Xiong2019} and the original method in \emph{Long et al.} \cite{long2017pde}. We specially develop a deep learning neural network in conjunction with sparse regression and Pareto analysis to deal with undefined terms. The \emph{mPDE-Net} method attempts to obtain a refined equivalent form of Vlasov equations based on machine learning to mimic the theoretical derivation of multiple moments by human beings to some extent.

The proposed \emph{mPDE-Net} has been tested for 1-D Landau damping as the proof-of-concept demonstration. A multi-moment fluid with closure model has been successfully learned from kinetic Vlasov data with a small amount of data sampling. Based on the learned multi-moment fluid PDEs, the time evolution of multiple physical parameters have been correctly predicted with initial and periodic boundary condition, which agrees with the kinetic landau damping rate well. This paper is organized as follows. Sec.~\ref{Methodology} describes methodology of \emph{mPDE-Net} and \emph{Time prediction}. Sec.~\ref{Simulation} describes simulation models of Landau damping of Langmuir waves, including kinetic model, learned fluid PDEs and multiple parameter prediction. The Landau damping of Langmuir waves learned by \emph{mPDE-Net} is compared with results from theoretical predictions and kinetic models. Finally, the conclusion and discussion will be summarized in Sec.~\ref{Conclusion}.  

\begin{figure*}
\centering
\includegraphics[width=0.85\textwidth]{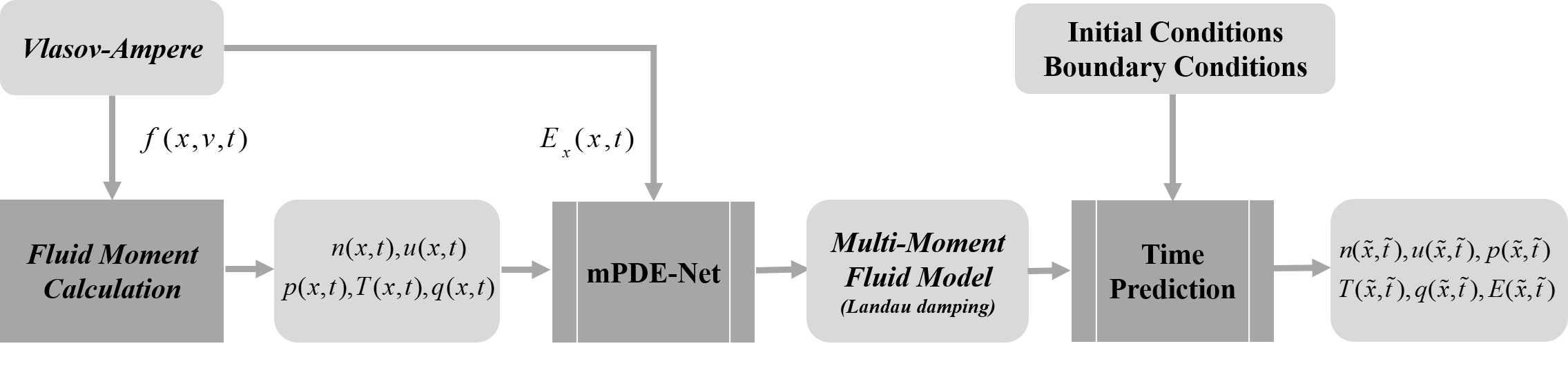}%
\caption{The schematic diagram of data-driven modeling and prediction architecture for Landau damping of Langmuir waves: learn the macroscopic multi-moment fluid model from kinetic data (numerical solution of Vlasov-Amp\'{e}re equations), and then make predictions based on learned fluid PDEs with initial and boundary conditions.}
\label{vlasov}
\end{figure*}

\section{\label{Methodology}Methodology}

The method proposed (\emph{mPDE-Net}) in this paper is an extension of the work of \emph{Xiong et al.} \cite{Xiong2019}. As we know, the multi-moment fluid model consists of three parts: the moment fluid equations, the field equations, and the fluid closure equations. In this work, we will focus on the one-dimensional (1-D) electrostatic case, and high-dimensional studies can be extended with the same method. The physical quantities involved in the 1-D electrostatic problem include density $ n(x,t) $, average velocity $ u(x,t) $, pressure $ p(x,t) $, temperature $ T(x,t) $, electric field $ E_{x}(x,t) $ and heat 
flux$ q(x,t) $. 
We assume that these physical quantities vary with time $t$ and the spatial coordinate $x$, which can be expressed in the form of a set of partial differential equations (PDEs),
\begin{equation}
\left\{ \begin{array}{l}
\vspace{1.5ex}
\frac{{\partial n}}{{\partial t}} = {f_n}(n,u,{E_x},\frac{{\partial n}}{{\partial x}},\frac{{\partial u}}{{\partial x}},\frac{{\partial {E_x}}}{{\partial x}})\\ \vspace{1.5ex}
\frac{{\partial u}}{{\partial t}} = {f_u}(n,u,p,{E_x},\frac{{\partial n}}{{\partial x}},\frac{{\partial u}}{{\partial x}},\frac{{\partial p}}{{\partial x}},\frac{{\partial {E_x}}}{{\partial x}})\\ \vspace{1.5ex}
\frac{{\partial p}}{{\partial t}} = {f_p}(n,u,p,q,{E_x},\frac{{\partial n}}{{\partial x}},\frac{{\partial u}}{{\partial x}},\frac{{\partial p}}{{\partial x}},\frac{{\partial q}}{{\partial x}},\frac{{\partial {E_x}}}{{\partial x}})\\\vspace{1.5ex}
\frac{{\partial T}}{{\partial t}} = {f_T}(n,u,T,q,{E_x},\frac{{\partial n}}{{\partial x}},\frac{{\partial u}}{{\partial x}},\frac{{\partial T}}{{\partial x}},\frac{{\partial q}}{{\partial x}},\frac{{\partial {E_x}}}{{\partial x}})\\\vspace{1.5ex}
\frac{{\partial {E_x}}}{{\partial t}} = {f_{{E_x}}}(n,u,{E_x},\frac{{\partial n}}{{\partial x}},\frac{{\partial u}}{{\partial x}},\frac{{\partial {E_x}}}{{\partial x}})\\\vspace{1.5ex}
q = {f_q}(n,u,T,\frac{{\partial n}}{{\partial x}},\frac{{\partial u}}{{\partial x}},\frac{{\partial T}}{{\partial x}})
\end{array} \right.
\label{eq1}
\end{equation}

where $f_{i}, i \in \{n, u, p, T, E_x, q\}$ are the nonlinear functions of all candidate terms in parentheses and their cross-terms. For the equations of $f_{i}$, where $i \in \{n, u, p, T\}$, the candidate terms include itself, higher-order moment, the external force, and their spatial differential terms. 
For $f_{{E_x}}$, the candidate terms include itself, fluid moment, and their spatial differential terms. 
For the third-order moment, $q$, the candidate terms in the closure equation ${f_{q}}$ include lower-order moments and their spatial differential terms. 
It is noteworthy that we will learn the PDEs of the Vlasov-Amp\'{e}re system in this study given that the Vlasov-Amp\'{e}re and Vlasov-Poisson systems are equivalent if the initial conditions of the Vlasov-Amp\'{e}re system satisfy Poisson's equation \cite{Xie2013}.

Fig. \ref{vlasov} depicts the overall framework of the proposed method, including spatiotemporal sampling, training, and time prediction. First of all, samples observed or calculated from spatio-temporal data in the small-scale kinetic region are fed into \emph{mPDE-Net}. Secondly, \emph{mPDE-Net} is activated to learn a set of PDEs which are capable of describing the fluid model with Landau damping. Finally, \emph{Time prediction} is adopted to predict multiple parameters based on learned multi-moment fluid models with closure.
\begin{figure*}
\centering
\includegraphics[width=0.85\textwidth]{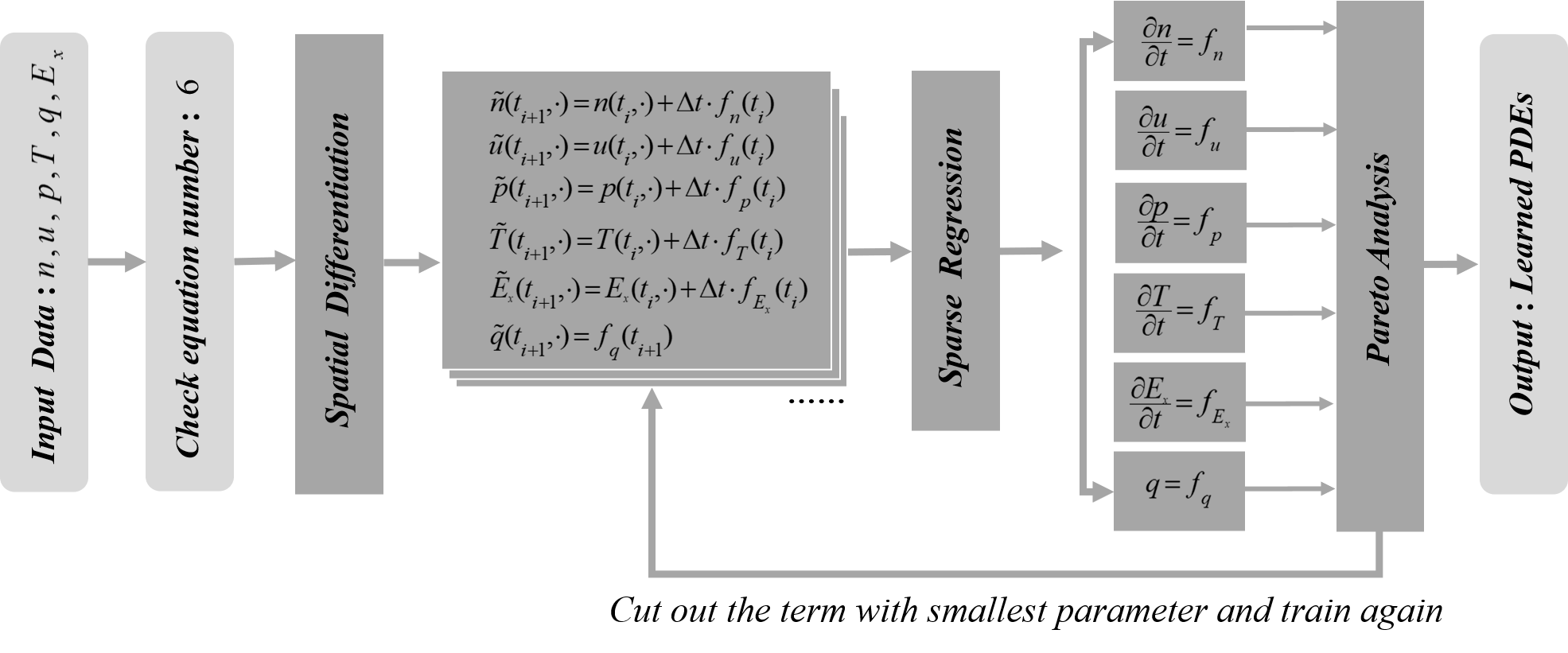}%
\caption{The schematic diagram of the \emph{mPDE-Net} architecture for data-driven modeling of the multi-moment fluid model with Landau damping}
\label{multi-moment fluid model-discover}
\end{figure*}

\subsection{mPDE-Net}
The \emph{mPDE-Net} architecture is shown in Fig. \ref{multi-moment fluid model-discover}.
The derivative of quantities with respect to space is calculated based on convolution \cite{cai2012image,dong2017image}. 
The central difference with the second-order accuracy is used to calculate the first-order and second-order spatial derivatives, respectively. 
The expressions of the convolution operator with the second order accuracy are as follows:
\begin{equation}
\begin{split}
h_{x}=\frac{1}{2}\begin{pmatrix}-1 & 0 & 1\end{pmatrix}\\
h_{xx}=\begin{pmatrix}1 & -2 & 1\end{pmatrix}
\end{split}
\label{eq}
\end{equation}
Then, the circular convolution (labeled as $\otimes$ ) between filters and function $f$ for instance is expressed as:
\begin{equation}
\begin{split}
h_{x}\otimes f\approx \delta_{x}\dfrac{\partial f}{\partial x}\\
h_{xx}\otimes f\approx \delta_{x}^{2}\dfrac{\partial^{2} f}{\partial x^{2}}
\end{split}
\label{eq}
\end{equation}
where $ \delta_{z}$ is the spatial size of $z$ direction of function $ u $.

Taking  velocity $u(x,t)$ as $f$ for instance, its spatial derivative with respect to $x$ can be calculated by the following formula:
\begin{equation}
\begin{split}
\dfrac{\partial u}{\partial x}\approx \dfrac{1}{2\delta_{x}}\begin{pmatrix}-1 & 0 & 1\end{pmatrix}\otimes u\\
\dfrac{\partial^{2} u}{\partial x^{2}}\approx \dfrac{1}{\delta_{x}^{2}}\begin{pmatrix}1 & -2 & 1\end{pmatrix}\otimes u
\end{split}
\label{eq_auax}
\end{equation}
Eq.\eqref{eq_auax} is also applicable for calculating spatial derivatives of other variables $u$, $p$, $T$, $E_{x}$, $q$.
In order to ensure accuracy, we do not calculate the spatial differentiation on the edges.

The time feedforward network is constructed by the forward Euler numerical scheme.
We define $ n(\cdot,t_{i}) $ as all spatial value of density $ n $ at $ t = t_{i} $, and $ \tilde{n}(\cdot,t_{i+1}) $ is the predicted value at the next time $t_{i+1}$.
According to the forward Euler method, the set of Eq \eqref{eq1} can be discretized for density $n$ as an example
\begin{equation}
\tilde{n}(\cdot,t_{i+1})=n(\cdot,t_{i})+(\Delta t)\cdot f_{n}(\cdot,t_{i})\\
\label{eq_time}
\end{equation}
where $ \tilde{n}(\cdot,t_{i+1}) $ is the approximation value at $ t_{i+1} $, and $\Delta t$ is the time step. 

The forward Euler $ \Delta t $ block is similar to \cite{long2017pde} as a layer of the neural network to advance variables based on equations in time. In principle, the structure of the $\Delta t $ block is an interpretation of Eq \eqref{eq_time}. The input of this network is possible candidate terms with $f$. The prior knowledge will help reduce the number of possible candidate functions associated with PDEs. The output of $ \Delta t $ block is the predicted value at the next time $\tilde{u}(\cdot,t_{i+1})$. In general, the entire network consists of several $\Delta t$ blocks by continuously connecting each block $ \Delta t $, where the number of $ \Delta t $ block is determined by certain convergence criteria.

In the neural network, the weights are shared in each layer, which serves as an output vector for the model. We define the weight vector $ \textbf{W} $ as 
\begin{equation}
\textbf{W}= [w_{1}, w_{2}, w_{3}, w_{4}, w_{5},...,w_{m}],
\label{eq6}
\end{equation}
where the non-zero weight component represents the term of the objective equation, and $m$ represents the number of candidate terms.
To reduce unnecessary candidates, we use a sparse regression method \cite{rudy2017data}, which uses $ L_{1} $ regularization. The cost function $L$ of the density $n$ equation is defined as 
\begin{equation}
L = \sum_{i=1}^N \parallel \tilde{n}(\cdot,t_{i})-n(\cdot,t_{i}) \parallel _{2}+\lambda_{n}\parallel \textbf{W} \parallel_{1}
\label{eq_loss_n}
\end{equation}
where $\lambda$ is the regularization parameter that affects the weight of the $ L_{1} $ regularization term, and $N$ is the time step. If a reasonable inversion result is not obtained initially with manually selected $\lambda$, we will automatically select the hyperparameter $\lambda$ using Grid Search (GS) optimization. 

Pareto analysis is a technique that can help identify top priority candidates of the concerned problem. The application of Pareto analysis is the final guarantee for the correct discovery of complex systems \cite{rudy2017data}. By cutting the term with the minimum parameter out and training again, we compare the later loss with the previous one. If loss reduces, then we repeat the process until loss increases. 

Based on the above modules, the construction procedure of the \emph{mPDE-Net} architecture can be described as following 

(1) Determine the number of required equations $m$.
If the number of categories of physical variables input by the network is $m_{0}$, then $m$ must satisfy $m \le {m_0}$. When $m \le {m_0}$, the system that needs to be learned is not closed; for $m = {m_0}$, the system is closed.

(2) Determine the order of the equation and the sampling data, calculate the spatial differentiation, and build the feedforward network. If equations are discretized in first-order, the first-order forward Euler is used to build the feedforward time network as shown in Eq. \eqref{eq_time}. If the equation is zeroth order, the  network is built directly according to the prior formula.

(3) Use sparse regression to optimize the weight $ \textbf{W} $.
Add $L_{1}$ regularization and use $Adam$ optimizer to train the network.

(4) Use Pareto analysis to segment the weight $ \textbf{W} $.
Based on each equation, Pareto analysis is used to optimize the weight $ \textbf{W} $ to achieve the most simplified form under the premise of ensuring accuracy.

(5) Repeat iteration until convergence, and finally output the identified equations.

\subsection{Time prediction}
Based on the learned multi-moment fluid model and initial conditions and periodic boundary conditions, the time prediction of the physical quantities can be written as follows,
\begin{equation}\label{pre_n}
\begin{aligned}
\tilde n( \cdot ,{t_{i + 1}}) = \Delta t \cdot {f_n} + n( \cdot ,{t_i})
\end{aligned}
\end{equation}

\begin{equation}\label{pre_u}
\tilde u( \cdot ,{t_{i + 1}}) = \Delta t \cdot {f_u} + u( \cdot ,{t_i})
\end{equation}

\begin{equation}\label{pre_p}
\tilde p( \cdot ,{t_{i + 1}}) = \Delta t \cdot {f_p} + p( \cdot ,{t_i})
\end{equation}

\begin{equation}\label{pre_T}
\tilde T( \cdot ,{t_{i + 1}}) = \Delta t \cdot {f_T} + T( \cdot ,{t_i})
\end{equation}

\begin{equation}\label{pre_Ex}
{{\tilde E}_x}( \cdot ,{t_{i + 1}}) = \Delta t \cdot {f_{E_x}} + {E_x}( \cdot ,{t_i})
\end{equation}

\begin{equation}\label{pre_q}
\tilde q( \cdot ,{t_{i + 1}}) = {f_q}
\end{equation}

\begin{figure}
\includegraphics[width=0.4\textwidth]{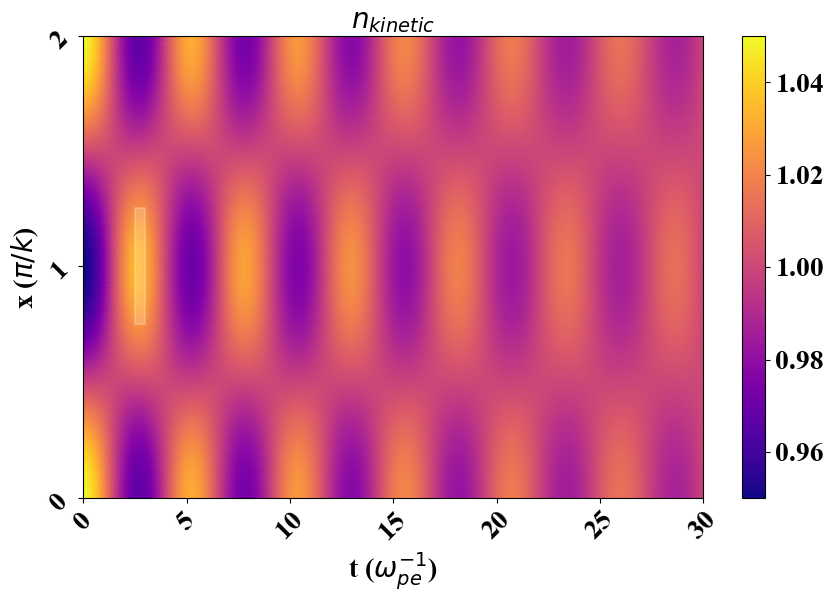}
\includegraphics[width=0.5\textwidth]{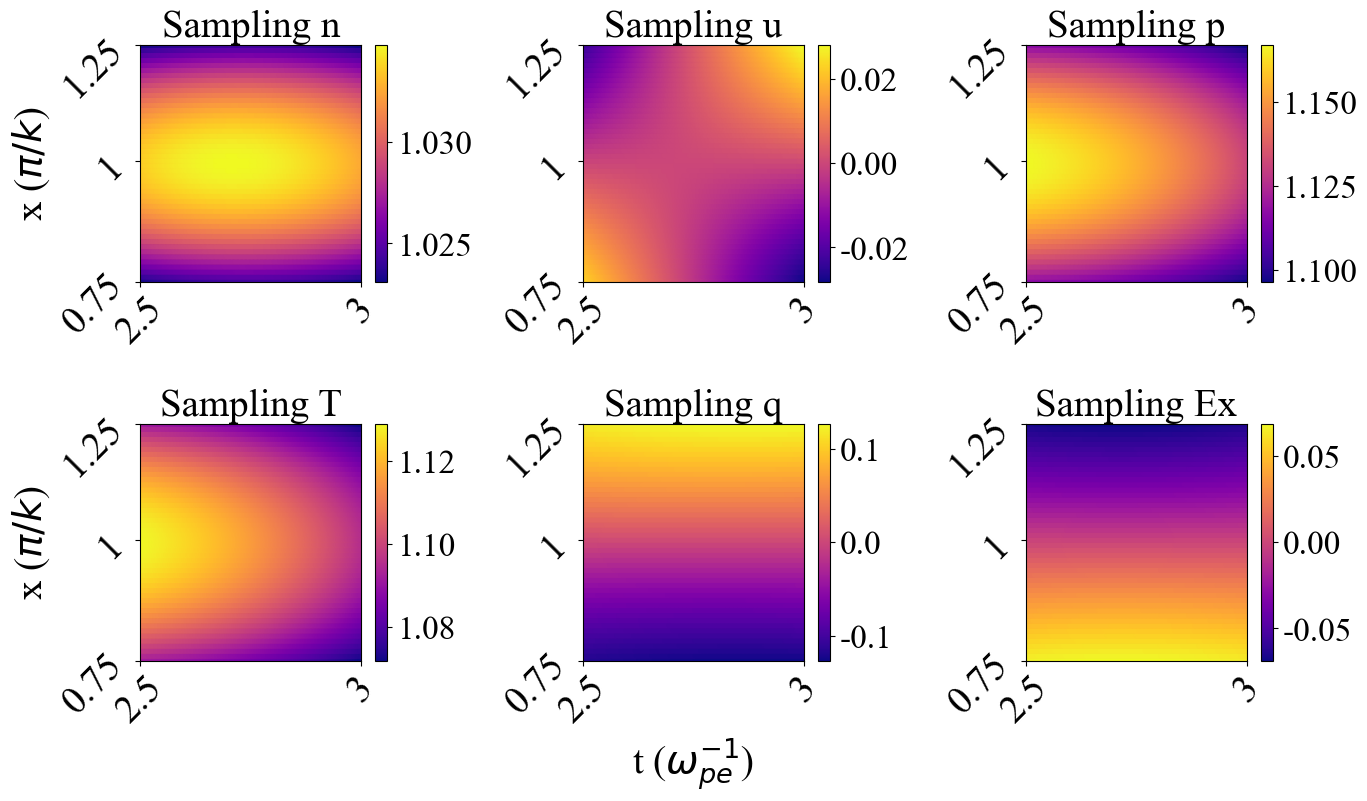}
\centering
\caption{The illustration of data sampling area for training from the kinetic model. The transparent box (white shaded) in the top figure depict the space-time sampling interval for $t = [2.5\omega_{pe}^{-1},3\omega_{pe}^{-1}]$ with $500$ points, $x = [0.75\pi /k,1.25\pi /k]$ with $48$ points. The bottom figure shows the zoomed sampling area for variables $n$, $u$, $p$, $T$, $q$, $E_x$ from kinetic data.}
\label{inputdata}
\end{figure}

\section{\label{Simulation}Simulation Test and Analysis}
\subsection{Kinetic model data for training}
For this test, we study the Landau damping of a Langmuir wave in 1X1V
by evolving the plasma phase-space distribution function $f\left(x,v,t\right)$. Landau
damping is one of the most fundamental plasma processes, and accurate
modeling this process is an important capability of any kinetic plasma
model. Numerically, we consider an immobile ion background with density
$n_{i}=n_{0}$ and a perturbed electron plasma with density 
\begin{equation}
n_{e}\left(x,t=0\right)=n_{0}\left(1+A\cos\left(kx\right)\right),
\end{equation}
where $k$ and $A$ are the perturbation wavenumber and amplitude,
respectively, and the electric field is perturbed to satisfy Gauss's
law. Unit normalization $\varepsilon_{0}=1$, $q_{e}=-1$, $m_{e}=1$,
$n_{0}=1$, and $T_{e}=1$ is used so that the electron plasma frequency
$\omega_{pe}=\sqrt{n_{e}q_{e}^{2}/m_{e}\varepsilon_{0}}=1$ and Debye
length $\lambda_{De}=\sqrt{\varepsilon_{0}T_{e}/n_{e}q_{e}^{2}}=1$.
The perturbation wavenumber $k$ and amplitude $A$ are set to $0.35\lambda_{De}^{-1}$
and 0.05. The periodic configuration domain of length $2\pi/k$ is
discretized by 64 cells. The velocity space domain $\left[-6v_{{\rm th},e},\ 6v_{{\rm th},e}\right]$
is discretized by 64 cells, where $v_{{\rm th},e}=\sqrt{T_{e}/m_{e}}=1$
is the background electron thermal velocity. Second-order serendipity
polynomial bases are used in space and the time integration uses the
strong-stability preserving Runge-Kutta scheme with a fixed time step
$\Delta t=0.001$ and stops at $t=30\omega_{pe}^{-1}$.

We obtain fluid quantities by calculating the moments of ${f}(x,v,t)$ in the velocity space, including electron density $n(x,t)$, average velocity $u(x,t)$, pressure $p(x,t)$, temperature $T(x,t)$ and heat flux $q(x,t)$.

\begin{equation}
n(x,t) = \int {f(x,v,t)dv}
\label{eq_n}
\end{equation}

\begin{equation}
u(x,t) = \frac{1}{{n(x,t)}}\int {vf(x,v,t)dv}
\label{eq_u}
\end{equation}

\begin{equation}
p(x,t) = m_e \int {{{(v - u(x,t))}^2}f(x,v,t)dv}
\label{eq_p}
\end{equation}

\begin{equation}
q(x,t) = m_e\int {{{(v - u(x,t))}^3}f(x,v,t)dv}
\label{eq_q}
\end{equation}

\begin{equation}
T(x,t) = \frac{{p(x,t)}}{{n(x,t)}}
\label{eq_T}
\end{equation}

The Vlasov-Amp\'{e}re equation is solved to obtain the kinetic data with the numerical code \texttt{Gkeyll} \cite{Juno2017}, including electric field ${E_x}(x,t)$, electron distribution function $f(x,v,t)$ and its moments in Eqs. (\ref{eq_n}-\ref{eq_q}). The electron temperature $T(x,t)$ is calculated from Eq. \eqref{eq_T}.

\subsection{Data-Driven Multi-Moment Fluid Simulation Results}

\renewcommand\arraystretch{2.5}
\begin{table*}
\caption{Comparison between ten-moment fluid model with gradient closure (Reference) and learned multi-moment fluid model by \emph{mPDE-Net} (Learned)}
\label{table-learn}
\centering
\begin{tabular}{m{5cm}<{\centering}|m{8cm}<{\centering}}
\hline
\hline

Reference & Learned \\ \hline
$\frac{{\partial n}}{{\partial t}} + u\frac{{\partial n}}{{\partial x}} + n\frac{{\partial u}}{{\partial x}} = 0$
& 
$\frac{{\partial n}}{{\partial t}} + 1.0008n\frac{{\partial u}}{{\partial x}} + 0.9972u\frac{{\partial n}}{{\partial x}} = 0$         \\

$\frac{{\partial u}}{{\partial t}} + u\frac{{\partial u}}{{\partial x}} + \frac{1}{{m_en}}\frac{{\partial p}}{{\partial x}} = \frac{q_e}{m_e}{E_x}$
& 
$\frac{{\partial u}}{{\partial t}} + 1.0035u\frac{{\partial u}}{{\partial x}} + 1.0038\frac{1}{{m_en}}\frac{{\partial p}}{{\partial x}} = 0.9978\frac{{{q_e}}}{{{m_e}}}{E_x}$        \\

$\frac{{\partial p}}{{\partial t}} + u\frac{{\partial p}}{{\partial x}} + 3p\frac{{\partial u}}{{\partial x}} + \frac{{\partial q}}{{\partial x}} = 0$
& 
$\frac{{\partial p}}{{\partial t}} + 0.9942u\frac{{\partial p}}{{\partial x}} + 2.9990p\frac{{\partial u}}{{\partial x}} + 1.0017\frac{{\partial q}}{{\partial x}} = 0$        \\

$\frac{{\partial T}}{{\partial t}} + 2T\frac{{\partial u}}{{\partial x}} + u\frac{{\partial T}}{{\partial x}} + \frac{1}{n}\frac{{\partial q}}{{\partial x}} = 0$
& 
$\frac{{\partial T}}{{\partial t}} + 1.9991T\frac{{\partial u}}{{\partial x}} + 0.9971u\frac{{\partial T}}{{\partial x}} + 1.0013\frac{1}{n}\frac{{\partial q}}{{\partial x}} = 0$       \\

$\frac{{\partial {E_x}}}{{\partial t}} + \frac{{{q_e}}}{{{\varepsilon _0}}}nu = 0$
&

$\frac{{\partial E_x}}{{\partial t}} + 1.0005\frac{{{q_e}}}{{{\varepsilon _0}}}nu = 0$         \\

$q = {\kappa}n\sqrt T \frac{{\partial T}}{{\partial x}}$
& 
$q =  - 4.2655n\sqrt T \frac{{\partial T}}{{\partial x}} 
+ \underbrace {\cancel{{\kappa _1}\frac{{{\partial ^2}T}}{{\partial {x^2}}}}}_0 + \underbrace {\cancel{{{\kappa _2}\frac{{\partial n}}{{\partial x}}\frac{{\partial T}}{{\partial x}}}}}_0
$
\\ \hline
\hline
\end{tabular}
\end{table*}

\begin{figure*}
\includegraphics[width=0.8\textwidth]{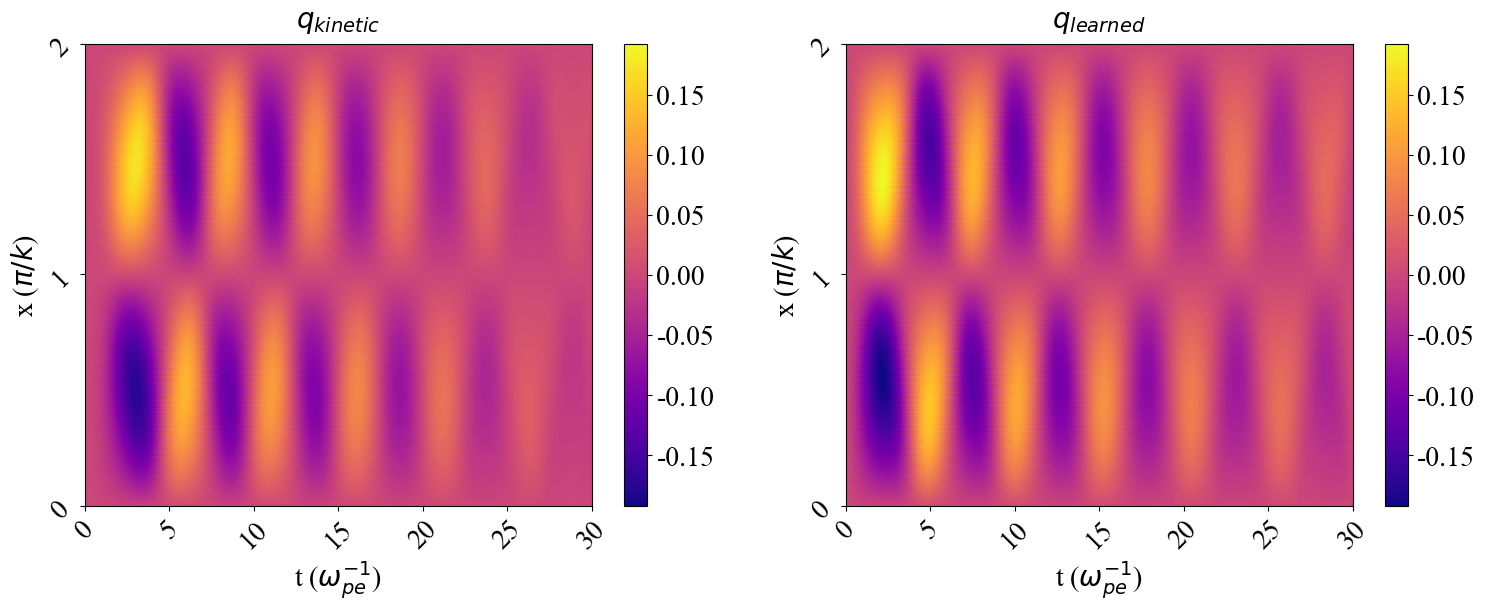}
\centering
\caption{The result of heat flux $q$ with learned closure form in Table \ref{table-learn} . The left figure is the heat flux from kinetic data and the right is the heat flux calculated by the learned heat flux closure in Table \ref{table-learn} with temperature and density data from kinetic simulations.
The time in $[0,30{\omega_{pe}^{-1}}]$ has $30000$ points, and the space in $[0,2\pi /k]$ has $192$ points.}
\label{q_clousre}
\end{figure*}

\begin{figure*}
\includegraphics[width=0.7\textwidth]{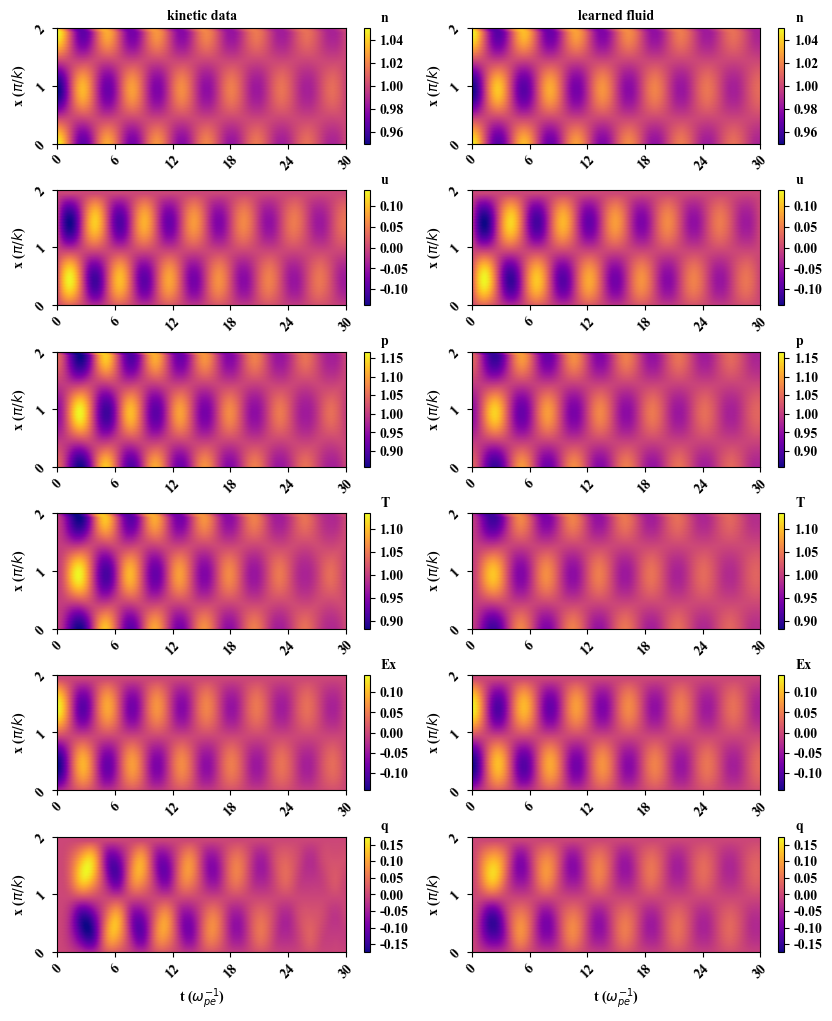}
\centering
\caption{Physical quantities calculated from the kinetic data (left) and predicted with the learned multi-moment fluid model (right). Each panel shows density $n$, velocity $u$, pressure $p$, temperature $T$, electric field $E_x$ and heat-flux $q$. The time is in $[0,30{\omega_{pe}^{-1}}]$ with $30000$ points and in space in $[0,2\pi /k]$ with $192$ points.}
\label{prediction}
\end{figure*}

\subsubsection{Constructing the data-Driven Multi-Moment Fluid  Model}
Based on our $1X1V$ kinetic simulation, we construct a machine learning-based neural network to learn a multi-moment fluid model from kinetic data as shown in Fig.\ref{multi-moment fluid model-discover}. 

We train the kinetic data $n$, $u$, $p$, $T$, $q$ and $E_x$ from \texttt{Gkeyll} in the sampling area by the \emph{mPDE-Net} algorithm to learn the form of the multi-moment fluid model equations. The sampling data is shown on the top panel of Fig. \ref{inputdata}, where the transparent box are temporal sampling interval $[2.5\omega_{pe}^{-1},3\omega_{pe}^{-1}]$ with $500$ points and spatial sampling interval $[0.75\pi /k,1.25\pi /k]$ with $48$ points. For clarity, the sampling scheme is zoomed out on the bottom panel of Fig. \ref{inputdata} for all variables $n, u, p, T, q, E_x$ as Fig. \ref{inputdata} .

According to previous studies on the heat flux closure \cite{Hammett1990,Jonathan2020}, it is generally believed that there exists a relationship between heat flux and temperature derivatives.
The original Hammett-Perkins closure \cite{Hammett1990} in Fourier space can be expressed as
\begin{equation}
\tilde{q}_{k}=-n_{\rm 0}\chi \frac{2^{1/2}v_{t}}{{|k|}} ik\tilde{T}_{k}
\end{equation}
where $\chi$ is dimensionless coefficients, $v_{t}$ is the electron thermal speed, $\tilde{T}_{k}=\tilde{p}-T_{\rm 0}\tilde{n}/n_{\rm 0}$ is the Fourier transformation of the perturbed temperature. 

As mentioned in \cite{Jonathan2020}, the resulting heat flux for single-mode can be expressed in configuration space as
\begin{equation}
q=-n\chi \frac{2^{1/2}v_{t}}{|k_{s}|}\frac{\partial{T}}{\partial{x}} 
\label{eq_p}
\end{equation}
where $k_{\rm s}$ is a constant or spatially varying parameter. 

It is important to note that the \emph{mPDE-Net} algorithm is capable of not only obtaining proper coefficients of terms for known PDEs but also removing trivial terms for unknown PDEs for the whole multiple-moment fluid system of equations. Table 1 shows the reference (left-column) and learned (right-column) multi-moment-Ampere equations, including the fluid closure. It should be noted that our training data $(n, u, p, T, E_x, q)$ is not constrained by the exact reference equations. Instead, we supplied a number of additional candidate terms to the neural network to allow it to identify the correct/expected terms based on the nonlinear physics contained in fully kinetic simulation data, eliminating the irrelevant candidates.
Particularly, for the heat flux $q$ closure equation,
we supplied candidate terms such as $\frac{{{\partial ^2}T}}{{\partial {x^2}}}$, $\frac{{\partial n}}{{\partial x}}\frac{{\partial T}}{{\partial x}}$, \emph{mPDE-Net} was able to discover the expected term $\kappa n \sqrt{T} \frac{\partial{T}}{\partial{x}}$ for $q$ with proper coefficients $\kappa$ and remove the redundant terms in Table \ref{table-learn}.

\begin{figure*}
\includegraphics[width=0.6\textwidth]{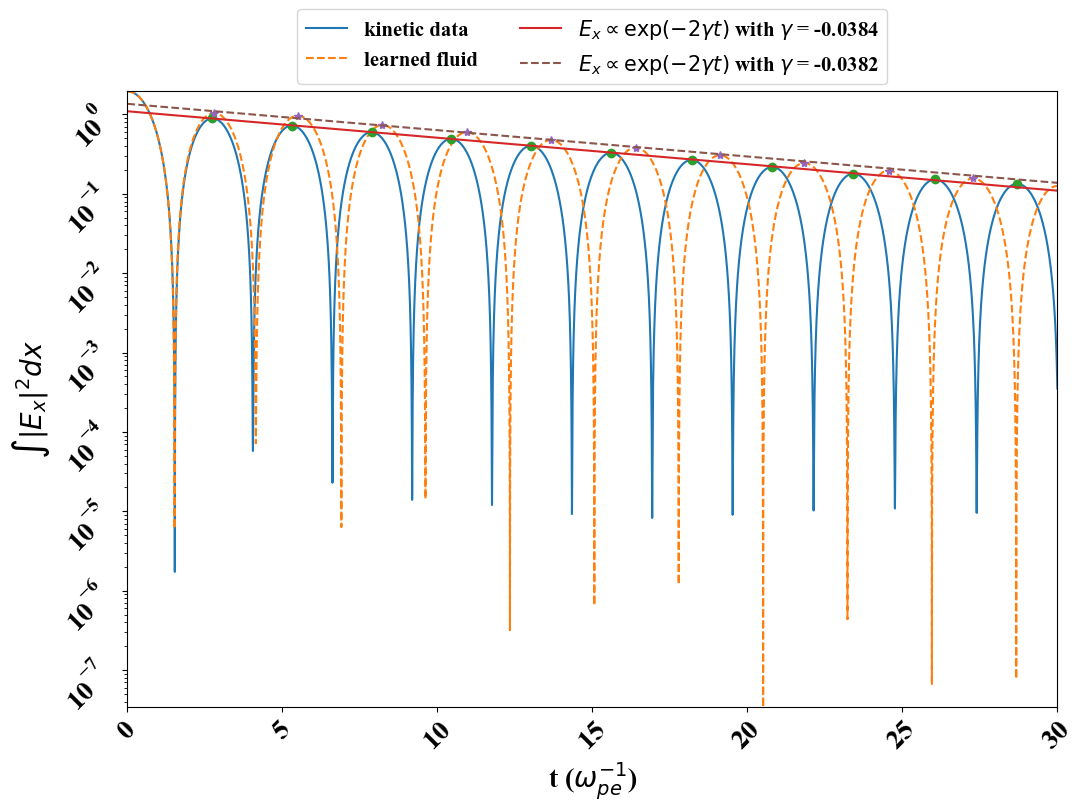}
\centering
\caption{Comparison of the evolution of electric field energy $\int {|E_x|^2} dx$ calculated from the kinetic data (blue curve) and predicted with the learned multi-moment fluid model (dashed yellow). The straight lines correspond to the exponential fit to the envelope of kinetic data (red) and learned fluid model (dashed purple) which provide the damping rate $\gamma /\omega_{pe} $.}
\label{pre_landau}
\end{figure*}

In order to evaluate performance, we define the relative error ${\delta}$ of each coefficient for the known fluid PDEs as follows:
\begin{equation}
\delta = \vert \dfrac{\tilde c-c}{c} \vert \times 100~\%
\end{equation}
where $\tilde c$ is the learned coefficient and $c$ is the theoretical coefficient. First of all, the relative coefficient errors of the continuity density equation are $0.08~\%$ and $0.28~\%$, respectively. Secondly, the relative coefficient errors of the velocity equation are $0.35~\%$, $0.38~\%$ and $0.22~\%$, respectively. Thirdly, the relative coefficient errors of the pressure equation are $0.58~\%$, $0.03~\%$ and $0.17~\%$, respectively. The relative coefficient errors of the temperature equation are $0.04~\%$, $0.29~\%$ and $0.13~\%$, respectively. Finally, the relative coefficient of the electric field equation is $0.05~\%$. 

Since the theoretical coefficient of the heat flux equation is unknown, in order to verify the heat flux closure equation in Table \ref{table-learn}, 
we compare the heat flux (right) based on learned fluid data with kinetic data (left) in Fig.\ref{q_clousre}. The comparison of the heat flux $q$ shows a similar result of heat flux between those calculated from kinetic data and learned from the neural network. 

\subsubsection{Time prediction}

\begin{figure*}
\includegraphics[width=0.8\textwidth]{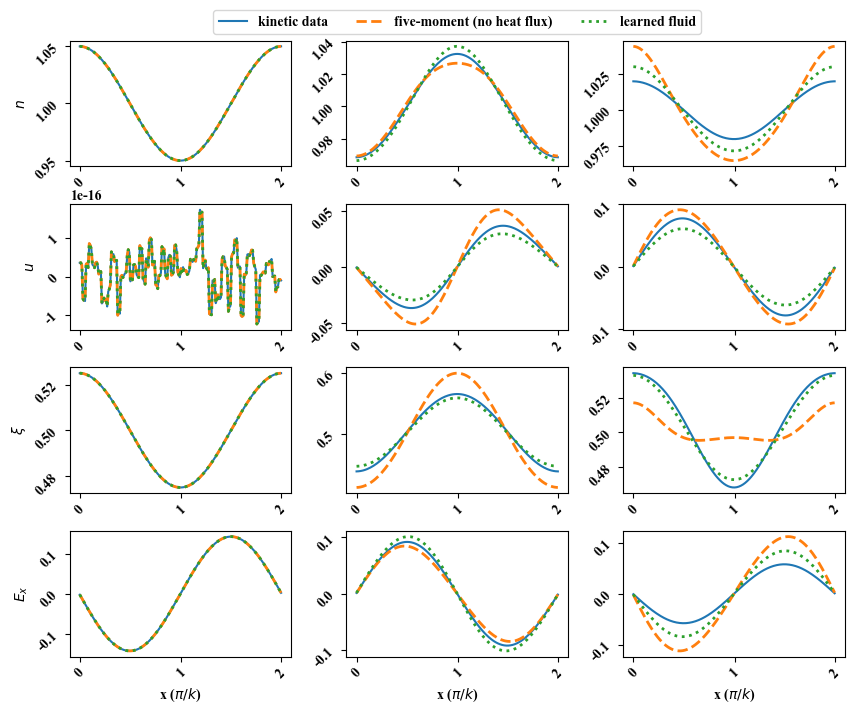}
\centering
\caption{Sample profiles of density $n$, velocity $u$, energy $\xi$ and electric field $E_x$ (from top to bottom) at $ t = 0, 3, 6$  (from left to right) in 1-D multi-moment fluid model, obtained from the kinetic equation, the five-moment equations, and learned multi-moment fluid model.}
\label{pre_compare}
\end{figure*}

\begin{figure*}
\includegraphics[width=0.8\textwidth]{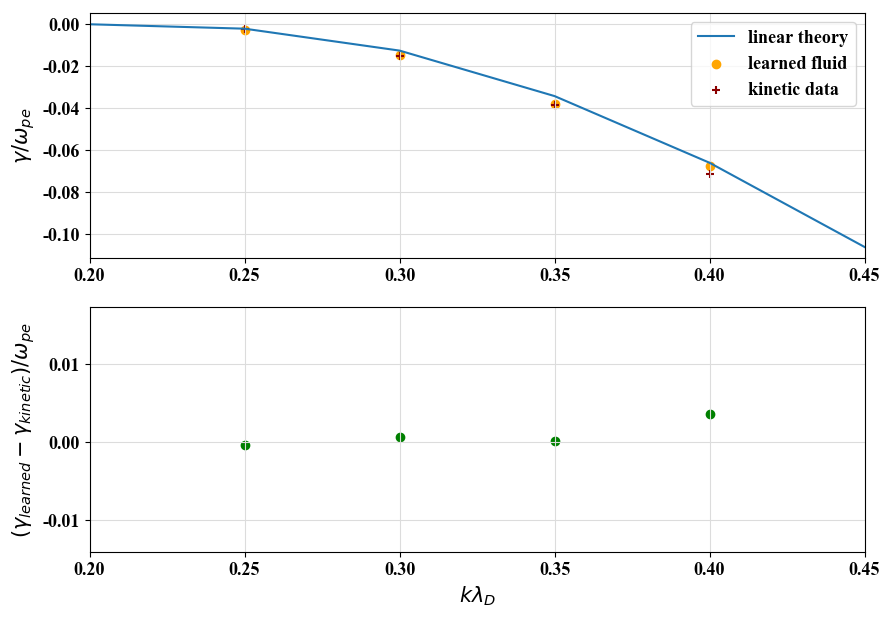}
\centering
\caption{The comparison of the Landau damping rates from the linear theory, the rates fitted to the kinetic simulation data and learned multiple-moment fluid equations. }
\label{Landau_with_k}
\end{figure*}

Now we have a multi-moment fluid model learned from kinetic data. 
Next, we will activate the \emph{Time Prediction} block to predict all physical quantities globally  based on the learned model with initial conditions.
The time prediction can verify whether the learned multi-moment fluid model is accurate, and it is also important for potential plasma applications. 

In our numerical scheme of time prediction, the time differentiation adopts the forward Euler and the space differentiation adopts the central difference with periodic boundary conditions. Based on the learned multi-moment fluid model (Table \ref{table-learn}) and the initial conditions, the time prediction of all six physical quantities are shown in Eq. ~\eqref{pre_n} - Eq. ~\eqref{pre_q} with periodic boundary conditions.

Fig.\ref{prediction} shows the prediction results in comparison with kinetic data, including density $n$, velocity $u$, pressure $p$, temperature $T$, electric field $E_x$ and heat-flux $q$.
Within the time interval $[0,30{\omega_{pe}^{-1}}]$, the prediction of all six physical quantities is in agreement with kinetic data. The errors gradually increase for a long time. One reason is that the coefficients of the learned multi-moment fluid equations contain small errors. The other one can be the numerical errors of the prediction schemes.

The most critical consequence of Landau damping is the exponential damping of the longitudinal charge waves.
In Fig.\ref{pre_landau}, we draw the time evolution of the total electric field energy predicted by the kinetic model versus that predicted by the learned fluid model.
The electric field energy bounces and decreases over time due to the damping. The local maximum values are linearly damped as expected. The learned model's bouncing frequency and damping rate both agree with the kinetic data.
A linear fitting of the maxima, in particular, yields the damping rates $\gamma=-0.0384\omega_{pe}^{-1}$ and $-0.0382\omega_{pe}^{-1}$ for the kinetic model and machine learning prediction, respectively, with a relative error of about 0.5 percent.
The bouncing frequency differs slightly between the two models.
We discovered that the learned fluid model result contains more less-damped modes with nontrivial contributions to the total energy after performing Fourier analysis on the time series.
These are the byproducts of the initial perturbation.
These undesirable "noise" modes are quickly damped in the fully kinetic model.
Because we assume only one mode dominates in constructing the closure in the learned fluid model, the learned closure does not provide sufficient damping to those noise modes that remain in the system and cause very small but persistent phase error.
Our numerical experiments demonstrated that using a smaller time step size effectively reduces phase errors.

Fig.\ref{pre_compare} compares the 1-D profile of density $n$, velocity $u$, energy $\xi$ and electric field $E_x$ profiles between the kinetic model, five-moment model (without heat flux $q$) and learned multi-moment fluid model (with heat flux $q$) at $t=0{\omega_{pe}^{-1}}$, $3{\omega_{pe}^{-1}}$, $6{\omega_{pe}^{-1}}$. The data from the learned multi-moment fluid model is consistent with the kinetic data, and shows improved accuracy than the well-known five-moment model.

In addition, we have also test \emph{mPDE-Net} for different wavenumber $k$. As shown in Fig. \ref{Landau_with_k}, the damping rates from the learned fluid model agree well with the kinetic data and the results from the linear dispersion relation solver. The absolute errors are less than $0.01$ with an optimized sampling scheme. Therefore, the learned multi-moment fluid model by \emph{mPDE-Net} can describe the kinetic Landau damping well.

\section{\label{Conclusion}Conclusion and Discussion}

In this paper, we proposed a data-driven network architecture to learn the multi-moment fluid model with Landau damping of Langmuir waves from kinetic data. A prediction scheme of the learned multi-moment fluid model has also been developed to achieve the prediction of Landau damping. The \emph{mPDE-Net} method has been verified for the 1-D electrostatic Vlasov-Amp\'{e}re system to well reproduce the Landau damping of the Langmuir wave. The Landau damping learned by \emph{mPDE-Net} has shown agreement with linear theory and kinetic simulations.
The \emph{mPDE-Net} architecture has the capability to achieve the physically interpretable PDEs set through machine learning.

In summary, our preliminary work adequately demonstrates the potential of learning and predicting plasma kinetic data by multi-moment fluid equations based on machine learning. Future work could consider a more general closure and extend the current architecture to high-dimensional plasma systems. The data-driven method of PDEs may pave the way to efficiently learning governing equations of complex, multi-scale, multi-physics systems.

\section*{Acknowledgments}
The authors from Fudan University were supported by the National Key Research and Development Program of China (2021YFA0717300) and China National Science Foundation (42074189). C.D. was partially supported by the U.S. Department of Energy under contract DE-AC02-09CH11466.


\end{document}